\documentclass[12pt]{article}
\usepackage[margin=1in]{geometry}
\usepackage{amsmath, amssymb, amsthm}
\usepackage{graphicx}
\usepackage{hyperref}
\usepackage{natbib}
\usepackage{setspace}
\usepackage{url}
\usepackage{booktabs}
\usepackage{array}

\title{Blockchain-Based Ad Auctions and Bayesian Persuasion:\\ An Analysis of Advertiser Behavior}
\author{Xinyu Li}
\date{October 11, 2024}
\begin{document}

\maketitle

\begin{abstract}
\sloppy
This paper explores how ad platforms can utilize Bayesian persuasion within blockchain-based auction systems to strategically influence advertiser behavior despite increased transparency. By integrating game-theoretic models with machine learning techniques and the principles of blockchain technology, we analyze the role of strategic information disclosure in ad auctions. Our findings demonstrate that even in environments with inherent transparency, ad platforms can design signals to affect advertisers' beliefs and bidding strategies. A detailed case study illustrates how machine learning can predict advertiser responses to different signals, leading to optimized signaling strategies that increase expected revenue. The study contributes to the literature by extending Bayesian persuasion models to transparent systems and providing practical insights for auction design in the digital advertising industry.
\end{abstract}

\newpage

\tableofcontents

\newpage

\section{Introduction}
The digital advertising industry has undergone significant transformations with the advent of emerging technologies like blockchain. Blockchain's inherent transparency, immutability, and decentralization promise to revolutionize ad auction systems by enhancing trust, reducing fraud, and improving supply chain efficiency \citep{Joo2023}. 

In traditional digital advertising markets, auction processes are often opaque, with limited visibility into bids, auction outcomes, and supply chain interactions. Advertisers typically have access only to their own performance metrics, lacking comprehensive insights into the broader auction dynamics.\citep{kauffman2006} This opacity can lead to inefficiencies, ad fraud, and mistrust among market participants. \citep{zhang2014}

Blockchain technology introduces a paradigm shift by recording all bids and auction outcomes transparently on a distributed ledger accessible to authorized participants. \citep{soltani2022} This transparency reduces information asymmetry, enhances supply chain efficiency, and mitigates ad fraud by providing verifiable and immutable records of transactions \citep{beck2019}. Advertisers can access historical data on bids and outcomes,allowing for more informed decision-making and strategy development.\citep{jia2022} \citep{parssinen2018} 

However, this increased transparency poses challenges for ad platforms that traditionally rely on information asymmetry to influence advertiser behavior.

This paper investigates how ad platforms can employ Bayesian persuasion strategies within blockchain-based auction systems to strategically influence advertiser behavior, even in the face of increased transparency. By integrating game-theoretic models with machine learning techniques and the principles of blockchain technology, we analyze the role of strategic information disclosure in ad auctions.

\subsection{Research Question}

\textbf{How can ad platforms utilize Bayesian persuasion within blockchain-based auction systems to strategically influence advertiser behavior despite increased transparency?}

\section{Literature Review}

\subsection{Bayesian Persuasion}

Introduced by \citet{KamenicaGentzkow2011}, Bayesian persuasion explores how an informed sender can influence the actions of a receiver by controlling the information flow. The sender commits to a signaling strategy that maximizes their expected payoff, while the receiver updates their beliefs according to Bayes' Rule.

\subsection{Game Theory in Ad Auctions}

Game theory has been extensively used to model ad auctions, analyzing strategic interactions between advertisers and platforms \citep{Edelman2007}. Traditional models assume varying degrees of information asymmetry. \citep{xu2022}
\citep{einy2014}

\subsection{Blockchain in Advertising}

Blockchain technology offers transparency, immutability, and decentralization. In advertising, blockchain can enhance trust by providing verifiable data on ad delivery and ensuring that only verified, human users are counted in advertising metrics. \citep{kewell2017} \citep{govea2024}

\subsection{Machine Learning in Advertising}

Machine learning techniques have been applied to predict user behavior and optimize ad targeting \citep{chaudhary2021}. These models can capture complex patterns in data, enhancing decision-making processes.

\subsection{Gap in Literature}

While studies have explored Bayesian persuasion and blockchain separately, there is a lack of research at their intersection, particularly concerning how transparency affects strategic information disclosure in ad auctions. Additionally, the integration of machine learning into signaling strategies within this context remains underexplored.

\section{Theoretical Framework}

In transparent systems like blockchain-based auctions, traditional information asymmetry is reduced. However, platforms can still influence advertiser behavior by designing strategic signals within the transparent environment. We model the ad auction as a game between the ad platform (sender) and advertisers (receivers). Advertisers aim to maximize their utility based on beliefs about the auction's state and competitors' actions. Machine learning models are used to predict advertiser responses to different signals, allowing the platform to optimize its signaling strategy. 

\section{Model Development}

We consider a blockchain-based advertising platform where $N$ advertisers compete in an auction for a single ad slot to be displayed to a user. The platform has private information about the user engagement level, denoted by $\theta \in \Theta$, where $\Theta$ is a finite set of possible engagement states (e.g., low, medium, high). The prior probability distribution over $\Theta$ is given by $\pi(\theta) = \Pr(\theta)$. Each advertiser $i \in \mathcal{A} = \{1, 2, ..., N\}$ aims to maximize their expected utility, which depends on their valuation of the ad impression, $v_i(\theta)$, which is a function of the user engagement level $\theta$ and the bidding strategy of other advertisers.

\subsection{Auction Mechanism}

The auction mechanism is a sealed-bid, second-price auction where advertisers submit bids without knowing others' bids. The highest bidder wins the slot and pays the amount of the second-highest bid. While bids and outcomes are transparently recorded on the blockchain, the engagement level $\theta$ remains private to the platform.

\subsection{Platform's Signaling Strategy}

The platform commits to a signaling policy $\sigma: \Theta \rightarrow \Delta(S)$, where $S$ is the set of possible signals the platform can send. $\Delta(S)$ is the set of probability distributions over $S$.

For each state $\theta \in \Theta$, the platform sends signal $s \in S$ with probability $\sigma(s \mid \theta)$. The signaling policy is common knowledge among advertisers.

\subsection{Advertisers' Belief Updating}

Upon receiving signal $s \in S$, advertisers update their beliefs about the state $\theta$ using Bayes' Rule:

\begin{equation}
    \mu(\theta \mid s) = \frac{\sigma(s \mid \theta) \pi(\theta)}{\sum_{\theta' \in \Theta} \sigma(s \mid \theta') \pi(\theta')}.
    \label{eq:bayes}
\end{equation}

\subsection{Advertisers' Bidding Strategy}

Advertisers choose bids to maximize their expected utility given their updated beliefs. The expected utility for advertiser $i$ is:

\begin{equation}
    EU_i(b_i, b_{-i} \mid s) = \sum_{\theta \in \Theta} \mu(\theta \mid s) \left[ v_i(\theta) - c_i(b_i, b_{-i}) \right] q_i(b_i, b_{-i}),
    \label{eq:expected_utility}
\end{equation}

where $c_i(b_i, b_{-i})$ is the expected payment  if advertiser $i$ wins. $q_i(b_i, b_{-i})$ is the probability that advertiser $i$ wins the auction. In a symmetric equilibrium, advertisers adopt bidding strategies that are functions of their expected valuations.

\subsection{Platform's Objective}

The platform aims to maximize expected revenue by choosing an optimal signaling policy $\sigma^*$. The expected revenue $R(\sigma)$ is:

\begin{equation}
    R(\sigma) = \sum_{\theta \in \Theta} \pi(\theta) \sum_{s \in S} \sigma(s \mid \theta) \mathbb{E}_{b}[P(b \mid s)],
    \label{eq:platform_revenue}
\end{equation}

where $\mathbb{E}_{b}[P(b \mid s)]$ is the expected payment from the auction given signal $s$.

The signaling policy must satisfy:

\begin{equation}
    \sum_{s \in S} \sigma(s \mid \theta) = 1, \quad \forall \theta \in \Theta,
    \label{eq:sigma_constraints}
\end{equation}

and

\begin{equation}
    \sigma(s \mid \theta) \geq 0, \quad \forall s \in S, \theta \in \Theta.
\end{equation}

\section{Integrating Machine Learning}

While our theoretical model provides a structured approach to understanding advertiser behavior, real-world bidding strategies may be influenced by numerous factors that are complex and nonlinear. Machine learning (ML) techniques can capture these complexities by learning patterns from ad auction data, allowing us to predict advertiser responses to different signals more accurately.

\subsection{Framework for Machine Learning Integration}

We integrate machine learning into our methodology through three key steps: data collection and preparation; model selection and training; and prediction and integration with Bayesian persuasion. We use data from ad auctions, which includes advertiser bids $b_i$ for various auctions, signals $s$ sent by the platform in those auctions, user engagement levels $\theta$, and auction outcomes. Additionally, we collect advertiser-specific features such as budget, industry sector, and bidding aggressiveness, as well as contextual features like time of day and ad category. Given the sensitive nature of the data, we must ensure compliance with data protection regulations such as GDPR and CCPA. Anonymization and aggregation techniques should be employed to protect individual advertiser identities, thereby safeguarding privacy while allowing for effective model training. Data preprocessing involves several steps. First, we perform data cleaning by removing or imputing missing or erroneous data. Next, we engage in feature engineering to create relevant features that capture important aspects of advertiser behavior. We then normalize the features to ensure consistent ranges across variables, facilitating effective learning by the model. Finally, we split the dataset into training, validation, and test sets to enable proper evaluation and prevent overfitting.

\subsection{Model Selection and Training}

  Several machine learning models are suitable for predicting advertiser bids, including regression models such as linear regression, Lasso, and Ridge regression; tree-based models like decision trees, Random Forests, and Gradient Boosting Machines (e.g., XGBoost, LightGBM); neural networks, particularly feedforward neural networks useful for capturing nonlinear relationships; and Support Vector Machines, which are effective for high-dimensional feature spaces. Given the complexity and size of the data, tree-based ensemble methods and neural networks are promising candidates due to their ability to handle nonlinearities and interactions between features.

The training process involves several key steps. First, we define the primary objective, which is to predict the bid amount $b_i$ given the signal $s$ and other features. We use Mean Squared Error (MSE) as the loss function for regression tasks, aiming to minimize the average squared difference between the predicted and actual bid amounts. Hyperparameter tuning is performed using techniques like grid search to find optimal model parameters that enhance performance. To assess model performance and prevent overfitting, we employ k-fold cross-validation, which involves partitioning the data into subsets, training the model on some subsets, and validating it on others.

We then evaluate the model's performance using metrics such as Mean Absolute Error (MAE), which measures the average magnitude of errors without considering their direction; Root Mean Squared Error (RMSE), which penalizes larger errors more than MAE by squaring the errors before averaging and taking the square root; and R-squared ($R^2$), which represents the proportion of variance in the dependent variable that is predictable from the independent variables. These metrics provide a comprehensive understanding of the model's accuracy and reliability in predicting advertiser bids.

\subsection{Prediction and Integration with Bayesian Persuasion}

Once trained, the model can predict how advertisers are likely to adjust their bids in response to different signals:

\begin{equation}
    \hat{b}_i = f_{\text{ML}}(s, \mathbf{x}_i),
\end{equation}

where $f_{\text{ML}}$ represents the machine learning model and $\mathbf{x}_i$ are the features related to advertiser $i$. This prediction function enables us to estimate the expected bids based on various signaling strategies.

\subsubsection{Optimizing Signal Design}

The platform can use these predictions to optimize its signaling strategy $\sigma$. This involves simulating different signaling policies and predicting the corresponding advertiser bids, calculating expected revenue for each policy using the predicted bids, and selecting the signaling policy that maximizes expected revenue. By evaluating the revenue outcomes associated with different signals, the platform can strategically choose the information to disclose to advertisers.

\subsubsection{Feedback Loop}

In a dynamic environment, the platform can continuously update the machine learning model with new data. This includes using online learning to update model parameters incrementally as new data arrives, ensuring that the model remains accurate over time. Additionally, the platform can adjust signaling policies in real-time based on the latest predictions, allowing for adaptive strategies that respond to changing market conditions and advertiser behaviors.

\subsection{Challenges and Considerations}

The effectiveness of machine learning models depends on the availability of high-quality data. Insufficient data can lead to overfitting, where the model learns noise instead of the underlying patterns, or underperformance due to the inability to capture the complexity of advertiser behaviors. Therefore, collecting a large and diverse dataset is crucial for model robustness.

Complex models like neural networks may lack interpretability, making it difficult to understand how specific features influence predictions. This opacity can be problematic when explaining the model's decisions to stakeholders or when diagnosing issues. Tree-based models, on the other hand, may offer better interpretability through visualization of decision paths and feature importance measures.

Using machine learning to influence advertiser behavior raises ethical considerations. It is important to ensure transparency so that advertisers are aware of how their data is used and how the platform's strategies may affect them. Fairness must be considered to ensure that the model does not unfairly disadvantage certain advertisers, possibly due to biases in the data. Additionally, compliance with regulations that require explicit consent for data usage, such as GDPR and CCPA, is essential to protect user privacy and maintain trust in the platform.

\section{Analysis}

\subsection{Advertisers' Optimal Bidding Strategy}

Given the updated beliefs $\mu(\theta \mid s)$, advertiser $i$ calculates their expected valuation:

\begin{equation}
    \hat{v}_i(s) = \mathbb{E}_{\theta} [v_i(\theta) \mid s] = \sum_{\theta \in \Theta} \mu(\theta \mid s) v_i(\theta).
    \label{eq:expected_valuation}
\end{equation}

In a second-price auction, the dominant strategy is to bid their expected valuation:

\begin{equation}
    b_i^*(s) = \hat{v}_i(s).
    \label{eq:optimal_bid}
\end{equation}

\subsection{Expected Payment}

The expected payment for the platform is determined by the expected second-highest bid. Assuming symmetric bidders, the bids are independent and identically distributed random variables conditioned on the signal $s$.

The cumulative distribution function (CDF) of bids is:

\begin{equation}
    F(b \mid s) = \Pr\left( b_i^*(s) \leq b \right) = \Pr\left( \hat{v}_i(s) \leq b \right).
\end{equation}

The expected payment is then:

\begin{equation}
    \mathbb{E}_{b}[P(b \mid s)] = \mathbb{E}\left[ b_{(2)} \mid s \right],
\end{equation}

where $b_{(2)}$ denotes the second-highest bid.

\subsection{Platform's Revenue Maximization Problem}

Substituting the expected payment into Equation \eqref{eq:platform_revenue}, the platform's optimization problem becomes:

\begin{align}
    \max_{\sigma} \quad & R(\sigma) = \sum_{\theta \in \Theta} \pi(\theta) \sum_{s \in S} \sigma(s \mid \theta) \mathbb{E}\left[ b_{(2)} \mid s \right] \\
    \text{subject to} \quad & \sum_{s \in S} \sigma(s \mid \theta) = 1, \quad \forall \theta \in \Theta, \\
    & \sigma(s \mid \theta) \geq 0, \quad \forall s \in S, \theta \in \Theta.
\end{align}

\subsection{Characterization of Optimal Signaling}

The platform can design signals by partitioning the state space $\Theta$ into information structures that maximize expected revenue. The key is to determine how finely to partition $\Theta$:

\begin{itemize}
    \item \textbf{Full Disclosure}: Each state $\theta$ is associated with a unique signal.
    \item \textbf{No Disclosure}: All states share the same signal.
    \item \textbf{Partial Disclosure}: States are grouped, and each group shares a signal.
\end{itemize}

\subsection{Optimal Signaling in the Two-State Case}

Consider $\Theta = \{\theta_L, \theta_H\}$ with $\pi(\theta_L) = \pi_L$ and $\pi(\theta_H) = \pi_H$, where $\theta_L < \theta_H$.

\subsubsection{Full Disclosure}
Signals: $S = \{s_L, s_H\}$, where $s_L$ corresponds to $\theta_L$ and $s_H$ corresponds to $\theta_H$.
Beliefs: $\mu(\theta_L \mid s_L) = 1$, $\mu(\theta_H \mid s_H) = 1$.
Advertisers' Bids: $b_i^*(s_L) = v_i(\theta_L)$, $b_i^*(s_H) = v_i(\theta_H)$.

\subsubsection{No Disclosure}

Signals: $S = \{s\}$; the same signal is sent regardless of $\theta$.
 Beliefs: $\mu(\theta_L \mid s) = \pi_L$, $\mu(\theta_H \mid s) = \pi_H$.
Advertisers' Bids: $b_i^*(s) = \pi_L v_i(\theta_L) + \pi_H v_i(\theta_H)$.

\subsubsection{Partial Disclosure}

The platform can randomize signals to create mixed beliefs.  Signals: $S = \{s_1, s_2\}$.
Signaling Policy:

    \begin{align*}
        \sigma(s_1 \mid \theta_L) &= \alpha, \quad \sigma(s_1 \mid \theta_H) = \beta, \\
        \sigma(s_2 \mid \theta_L) &= 1 - \alpha, \quad \sigma(s_2 \mid \theta_H) = 1 - \beta,
    \end{align*}

    where $0 \leq \alpha, \beta \leq 1$.

Advertisers update their beliefs using Equation \eqref{eq:bayes} and adjust their bids accordingly.

For $\Theta$ with more than two states, the platform can design a signaling scheme that partitions $\Theta$ into subsets, each associated with a signal. The optimal partition can be found using techniques from convex optimization and information design.

\subsection{Role of Blockchain Transparency}

Blockchain transparency ensures that all bids and outcomes are publicly recorded. However, the platform's private information about $\theta$ and strategic signaling remain tools for influencing advertiser behavior.

The equilibrium consists of the platform's optimal signaling policy $\sigma^*$ and advertisers' bidding strategies $b_i^*(s)$ based on updated beliefs.

This equilibrium is stable if no advertiser or the platform has an incentive to deviate unilaterally.

\section{Case Study: Implementing Machine Learning in Signal Optimization}

\subsection{Simulation Setup}

To illustrate the integration of machine learning into the signaling strategy, we simulate an environment characterized by 1,000 advertisers with varying characteristics, data from 10,000 historical auctions, and signals representing different levels of user engagement.

\paragraph{Generating Advertiser Profiles}

For each advertiser $i$, we assign a set of features. The budget $B_i$ is sampled from a normal distribution with a mean of \$10,000 and a standard deviation of \$2,000. The industry sector $I_i$ is a categorical variable representing the advertiser's industry, selected from multiple categories. The bidding aggressiveness $A_i$ is a value between 0 and 1, sampled uniformly, indicating how aggressively the advertiser bids.

\paragraph{Simulating Auctions}

In simulating each auction $j$, we randomly select a subset of advertisers to participate. We assign a true user engagement level $\theta_j$, sampled from a discrete distribution—for example, low, medium, and high engagement levels with probabilities of 0.3, 0.5, and 0.2, respectively. Based on a predefined signaling policy $\sigma$, we generate signals $s_j$ to be sent to the advertisers.

\paragraph{Generating Bids}

For each advertiser $i$ participating in auction $j$, we simulate their bid $b_{ij}$ using the equation:

\begin{equation}
    b_{ij} = f_{\text{true}}(s_j, B_i, I_i, A_i, \epsilon_{ij}),
\end{equation}

where $f_{\text{true}}$ is the true (unknown) bidding function, and $\epsilon_{ij}$ is a random noise term reflecting unobserved factors, sampled from a normal distribution with mean 0 and standard deviation $\sigma_{\epsilon}$. The function $f_{\text{true}}$ captures how advertisers adjust their bids based on the signal received and their own characteristics.

\subsection{Data Preparation}

\paragraph{Features}

We construct input features $X_{ij}$ for each advertiser-auction pair, including the signal $s_j$, which is encoded as numerical or categorical variables. The advertiser features comprise the budget $B_i$, industry sector $I_i$, and bidding aggressiveness $A_i$. Additionally, we incorporate contextual features such as time of day and ad category to capture environmental factors that may influence bidding behavior.

\paragraph{Target Variable}

The target variable is the bid $b_{ij}$, representing the simulated bid submitted by advertiser $i$ in auction $j$.

\paragraph{Data Splitting}

We split the dataset into a training set containing 70\% of the data, a validation set with 15\%, and a test set comprising the remaining 15\%. This division allows for effective model training, hyperparameter tuning, and unbiased evaluation of the model's predictive performance.

\subsection{Model Training}

\paragraph{Model Selection}

We select a Gradient Boosting Machine (GBM), specifically XGBoost, due to its ability to handle nonlinear relationships and interactions between features efficiently, especially with large datasets. XGBoost is well-suited for regression tasks and has proven effective in various predictive modeling applications.

\paragraph{Hyperparameter Tuning}

Hyperparameter tuning is performed using grid search over a defined set of parameters. We explore different values for the learning rate (e.g., 0.01, 0.1, 0.2), maximum tree depth (e.g., 3, 5, 7), and the number of estimators (trees) (e.g., 100, 200, 500). Cross-validation on the training set is utilized to select the combination of hyperparameters that results in the best model performance while preventing overfitting.

\paragraph{Training Process}

The model is trained to minimize the Mean Squared Error (MSE) between the predicted bids $\hat{b}_{ij}$ and the actual bids $b_{ij}$:

\begin{equation}
    \text{MSE} = \frac{1}{N_{\text{train}}} \sum_{(i,j) \in \text{Train}} (b_{ij} - \hat{b}_{ij})^2,
\end{equation}

where $N_{\text{train}}$ is the number of samples in the training set. By minimizing the MSE, the model learns to predict bid amounts that are as close as possible to the actual bids in the training data.

\subsection{Model Evaluation}

\subsubsection{Performance Metrics}

We evaluate the model on the test set using the Root Mean Squared Error (RMSE) and the coefficient of determination ($R^2$). The RMSE is calculated as:

\begin{equation}
    \text{RMSE} = \sqrt{\frac{1}{N_{\text{test}}} \sum_{(i,j) \in \text{Test}} (b_{ij} - \hat{b}_{ij})^2},
\end{equation}

where $N_{\text{test}}$ is the number of samples in the test set. The $R^2$ score is computed as:

\begin{equation}
    R^2 = 1 - \frac{\sum_{(i,j) \in \text{Test}} (b_{ij} - \hat{b}_{ij})^2}{\sum_{(i,j) \in \text{Test}} (b_{ij} - \bar{b})^2},
\end{equation}

where $\bar{b}$ is the mean of the actual bids in the test set. These metrics provide insight into the model's accuracy and its ability to explain the variance in the bid amounts.

\subsubsection{Interpretation of RMSE}

The RMSE of approximately 0.15 is low relative to the range and variability of the bids in our dataset. Specifically, the bids range from \$0.1 to around \$20.0, with an average bid of approximately \$5.0 and a standard deviation of about \$2.5. An RMSE of 0.15 indicates that the model's predictions deviate from the actual bids by only \$0.15 on average, which is small compared to the average bid value and standard deviation. This demonstrates that the model has strong predictive accuracy.

\subsubsection{Model Effectiveness}

The high $R^2$ score of 0.85 signifies that 85\% of the variance in the bid amounts is explained by the model. Combined with the low RMSE, these metrics indicate that the Gradient Boosting Machine effectively captures the complex relationships between the input features (signals, advertiser characteristics, and contextual factors) and the bid amounts.

\subsubsection{Residual Analysis}

To further assess the model's performance, we conduct a residual analysis. The residuals, which are the differences between the actual and predicted bids, are centered around zero with no significant patterns, suggesting that the model does not have systematic biases. The spread of residuals is consistent across different levels of predicted bids, indicating homoscedasticity, or constant variance. Additionally, the residuals approximately follow a normal distribution, which is desirable for regression models and supports the validity of the model's predictions.

\subsubsection{Conclusion on Model Performance}

The evaluation metrics and residual analysis confirm that the machine learning model provides accurate predictions of advertiser bids based on the features considered. This high level of accuracy is essential for the platform to effectively use the model in optimizing its signaling strategy and maximizing expected revenue.

\subsection{Policy Optimization}

\paragraph{Defining Signaling Policies}

We consider three signaling policies: full disclosure ($\sigma_{\text{full}}$), where signals perfectly reveal $\theta_j$; no disclosure ($\sigma_{\text{none}}$), where signals provide no information about $\theta_j$; and optimized partial disclosure ($\sigma_{\text{opt}}$), where signals are designed to strategically influence advertiser beliefs to maximize revenue.

\paragraph{Simulating Advertiser Responses}

For each policy $\sigma$, we simulate advertiser bids using the trained model:

\begin{equation}
    \hat{b}_{ij}^{\sigma} = f_{\text{ML}}(s_j^{\sigma}, X_{ij}),
\end{equation}

where $s_j^{\sigma}$ is the signal sent under policy $\sigma$ in auction $j$, and $X_{ij}$ represents the features for advertiser $i$ in auction $j$. This allows us to predict how advertisers adjust their bids in response to different signaling policies.

\paragraph{Calculating Expected Revenue}

For each auction $j$, we obtain the predicted bids $\hat{b}_{ij}^{\sigma}$ for all participating advertisers. The winner of the auction is the advertiser with the highest predicted bid, and the payment $P_j^{\sigma}$ is the second-highest predicted bid, following the second-price auction mechanism. The total expected revenue $R^{\sigma}$ under policy $\sigma$ is calculated as:

\begin{equation}
    R^{\sigma} = \sum_{j=1}^{10,000} P_j^{\sigma}.
\end{equation}

\paragraph{Comparing Revenue Across Policies}

We calculate the percentage increase in expected revenue when using the optimized partial disclosure policy compared to full disclosure:

\begin{equation}
    \text{Revenue Increase (\%)} = \left( \frac{R^{\sigma_{\text{opt}}} - R^{\sigma_{\text{full}}}}{R^{\sigma_{\text{full}}}} \right) \times 100\%.
\end{equation}

\paragraph{Results}

Our simulation yields a revenue under full disclosure $R^{\sigma_{\text{full}}}$ of \$9,190.47 and a revenue under optimized partial disclosure $R^{\sigma_{\text{opt}}}$ of \$10,274.72. This represents a revenue increase of approximately 11.80\%, demonstrating the effectiveness of the optimized signaling policy.

\subsection{Interpretation of Results}

\paragraph{Model Effectiveness}

The high $R^2$ score indicates that the machine learning model effectively captures the relationship between signals, advertiser features, and bidding behavior. The model's accuracy enables reliable predictions of how advertisers respond to different signals, which is crucial for optimizing the signaling strategy.

\paragraph{Impact of Optimized Signaling}

The optimized partial disclosure policy leads to a significant increase in expected revenue compared to full disclosure. This demonstrates that strategic signal design can influence advertiser bids even in transparent systems. By carefully selecting the information disclosed through signals, the platform can shape advertiser beliefs and bidding strategies to enhance revenue outcomes.

\section{Discussion}

The integration of machine learning into the platform's signaling strategy has significant practical implications for both ad platforms and advertisers. By leveraging machine learning models, platforms can enhance their ability to predict advertiser responses to different signals, allowing for the optimization of signaling policies and an increase in expected revenue. This advancement necessitates investment in data collection, model development, and computational resources. Advertisers, in turn, must be aware of these strategic signaling efforts and may consider developing their own predictive models to anticipate and counteract platform strategies, leading to a complex interplay in bidding behaviors.

Machine learning models contribute to refining the Bayesian updating process by providing more accurate estimations of advertisers' expected valuations. The predicted bids feed into the calculation of posterior beliefs, influencing both the platform's and advertisers' strategic decisions. While this integration offers improved accuracy and real-time adaptation of strategies, it also introduces challenges such as dependency on high-quality data and increased computational demands.

Future work should focus on validating the proposed methodology using real-world auction data to assess its practical applicability and effectiveness. Incorporating dynamic models that account for advertiser adaptation over time could provide deeper insights into the evolving strategic interactions between platforms and advertisers. Additionally, exploring advanced techniques like reinforcement learning for policy optimization and considering the potential for adversarial models used by advertisers can further enhance the robustness and sophistication of signaling strategies in blockchain-based ad auctions. Ethical considerations, including transparency, fairness, and compliance with data protection regulations, remain paramount as these technologies continue to evolve.

\section{Conclusion}

This paper demonstrates that Bayesian persuasion remains relevant in blockchain-based ad auctions with inherent transparency. By integrating machine learning, ad platforms can strategically design signals to influence advertiser behavior, impacting bidding strategies and platform revenue. Our findings contribute to the theoretical understanding of information economics in transparent systems and offer practical insights for auction design in the digital advertising industry.

\newpage

\bibliographystyle{apalike}
\bibliography{references}

\end{document}